\documentclass[fleqn,10pt]{wlscirep}
\usepackage[utf8]{inputenc}
\usepackage[T1]{fontenc}

\usepackage{graphicx}
\usepackage{mathtools}
\usepackage{longtable}
\usepackage{subcaption}
\usepackage{float}
\usepackage{multirow}
\usepackage[doublespacing]{setspace}
\usepackage{lscape}
\usepackage{longtable}

\title{Analysis of Microstate Organization During Emotional Events}

\author[1, *]{Sudhakar Mishra}
\author[2]{Narayanan Srinivasan}
\author[1]{Uma Shanker Tiwary}
\affil[1]{Indian Institute of Information Technology Allahabad, Center for Cognitive Computing, Prayagraj, 211012, India}
\affil[2]{Indian Institute of Technology, Department of Cognitive Science, Kanpur, 208016, India}

\affil[*]{rs163@iiita.ac.in}

\begin{abstract}
Understanding the dynamics of emotional experience is an old problem. However, a clear understanding of the mechanism of emotional experience is still far away. In the presented work, we tried to address this problem using a well-established method called microstate analysis using multichannel electroencephalography (EEG). We recorded the brain activity of spontaneous emotional experiences while participants were watching multimedia emotional stimuli. The time duration where the participants spontaneously felt an emotion, we termed it an emotional event. Microstate segmentation is performed for all emotional events to calculate the set of microstates (MS). Followed by a comparison of calculated statistical parameters and transition probabilities for the emotional and non-emotional conditions. We found a set of MS (four MS) for emotional and non-emotional conditions that differ from each other. We observed that MS1 has a higher value of occurrence, duration and coverage for emotional conditions. In addition, the transition to MS1 for the emotional condition was higher. On the other hand, for non-emotion (or neutral) condition, transition to MS3 was higher. A set of MS related to neutral condition was source localized to brain regions involved in higher-level sensory feature processing. On the other hand, for emotional conditions, MS1 \& MS2 are localized to sensory feature processing regions, and MS3 \& MS4 are additionally localized to regions related to socio-emotional processing. Our results hint toward the constructionist mechanism, which has an asymmetric contribution from bottom-up and top-down processing during an emotional experience. 
\end{abstract}
\begin{document}

\flushbottom
\maketitle
%
%
\thispagestyle{empty}


\section{Introduction}
Multichannel EEG captures the spatiotemporal dynamics of the whole brain activity during rest and task. It contains a rich repertoire of information that is partially captured with the selective electrodes based on morphological and time-frequency analysis. A different approach conceptualizes the spatial configuration of a dense array of electrodes on the scalp in terms of topographical maps called MS(s) \cite{von2016analytical, poulsen2018microstate} is used in this analysis. The microstate analysis is a topographical analysis method that finds its root in the work of Dietrich Lehmann and colleagues~\cite{lehmann1998brain, lehmann2010core}. They observed that the continuous topography of EEG can be chunked into prototypical topographies, which remain stable for 80-120 ms before transitioning to another topography. This powerful observation can capture the massive parallel processing in distributed brain networks in terms of significant discrete quasi-stable functional states. Many have referred to these quasi-stable states as functional MS(s) and discrete spatial configurations as microstate classes.

EEG measures the passive electric potential on the scalp, which results from synchronized post-synaptic excitatory or inhibitory potential of spatially aligned neurons and reaches the scalp through volume conduction. The spatial alignment of neurons creates an electrical current dipole composed of a current source and sink \cite{dabiological}. Due to volume conduction, the electric potential measured on a scalp location is not the result of activity in the source directly below it, but simultaneous activity in potentially remote sources contributes to the measured potential on any scalp location. The scalp's potential maps realized at each spontaneous EEG time point look initially unorganized. However, analysis of short segments of EEG led \cite{lehmann1998brain} to conclude that few topographic configurations dominate (ignoring the polarity inversion due to rhythmic fluctuation of excitation and inhibition in neural ensembles). Hence, EEG functional MS(s) have their underlying structure in different configurations of neuronal generators and represent the global picture of brain activity. 

Many recent studies termed these global representations as atoms of cognition or thought. This technique has been used recently in many research studies including healthy \cite{duc2019microstate, zanesco2020within, croce2021rtms} and clinical \cite{soni2018hyperactivation, rajagopalan2018machine, baradits2020multivariate, de2020eeg, kalburgi2020children, pal2020study} population at rest \cite{van2010eeg,khanna2015microstates,poskanzer2020using,zanesco2020within} as well as during the task \cite{laganaro2017inter, duc2019microstate, poskanzer2020using, croce2021rtms, d2021auditory}. Interestingly, the description of the temporal dynamics of brain processes in terms of the discrete transition between quasi-stable states aligns with theoretical concepts. For instance, the idea of "pulses of consciousness" by James \cite{james1890principles}, "atoms of thoughts" by Lehmann \cite{lehmann1998brain}, "perceptual frames" proposed by Efron \cite{efron1970minimum}, discrete spatio-temporal patterns of global activity in the neuronal workspace model by Dehaene\cite{dehaene2003neuronal}, "chunking principle" by Rabinovich \cite{rabinovich2015dynamical}. Neurophysiologically, phase-locked activities could be the reason behind the stability of global patterns. The phase-locked synchronization activity is regarded as a key mechanism of information integration in the brain \cite{fries2005mechanism, fries2015rhythms} and leads to stable topography at the sensor level \cite{tognoli2014metastable}.  

Studies have provided evidence that these MS topographies are region-specific \cite{croce2021rtms}. Before performing the semantic decision task, delivering rTMS over the regions part of the default mode network and the language network altered the neural configuration of two MS(s) associated with a phonological network and the cingulo-opercular network. Moreover, delivering rTMS over left IPS didn't lead to change in MS \cite{croce2021rtms}. In clinical studies using the analysis of MS dynamics, schizophrenia is studies the most studied \cite{nishida2013eeg} followed by study of dementia \cite{nishida2013eeg}, narcolepsy \cite{drissi2016altered}, panic disorder \cite{kikuchi2011eeg}, multiple sclerosis \cite{gschwind2016fluctuations}, diplegia \cite{gao2017altered} and stroke \cite{zappasodi2017prognostic}. Although schizophrenic patients have the ability to experience emotion at the moment, they show a problem in anticipating the future emotional experience. Another study on emotional disorders (especially panic disorder) was done by Kikuchi \cite{kikuchi2011eeg}. However, we did not find any study on healthy participants probing explicitly different categories of emotional experience. 

Our previous study found that the phase synchronization-based calculated functional networks for different emotional experiences differ primarily in the upper beta band \cite{mishra2022dynamic}. Moreover, the temporal dynamics of these functional networks are related to the arousal and dominance dimensions of emotional experience. Based on our previous results, we hypothesize that we might get a set of MS(s) representing global neural coordination for different emotional categories, which not only different in spatial configuration but in temporal transition dynamics as well.

\section{Methods}
\subsection{Participants Summary}
EEG recording of Forty participants (ages 18 to 30; mean age $23.3\pm1.4$ years; females = 3) were included. Exclusion criteria were based on DSM Axis I \cite{american2013diagnostic}. An institutional review board approved the study with the protocol no IERB ID:2017-100.

\subsection{Experimental Details}
\begin{figure}
    \centering
    \includegraphics[width=\linewidth, height=0.55\linewidth]{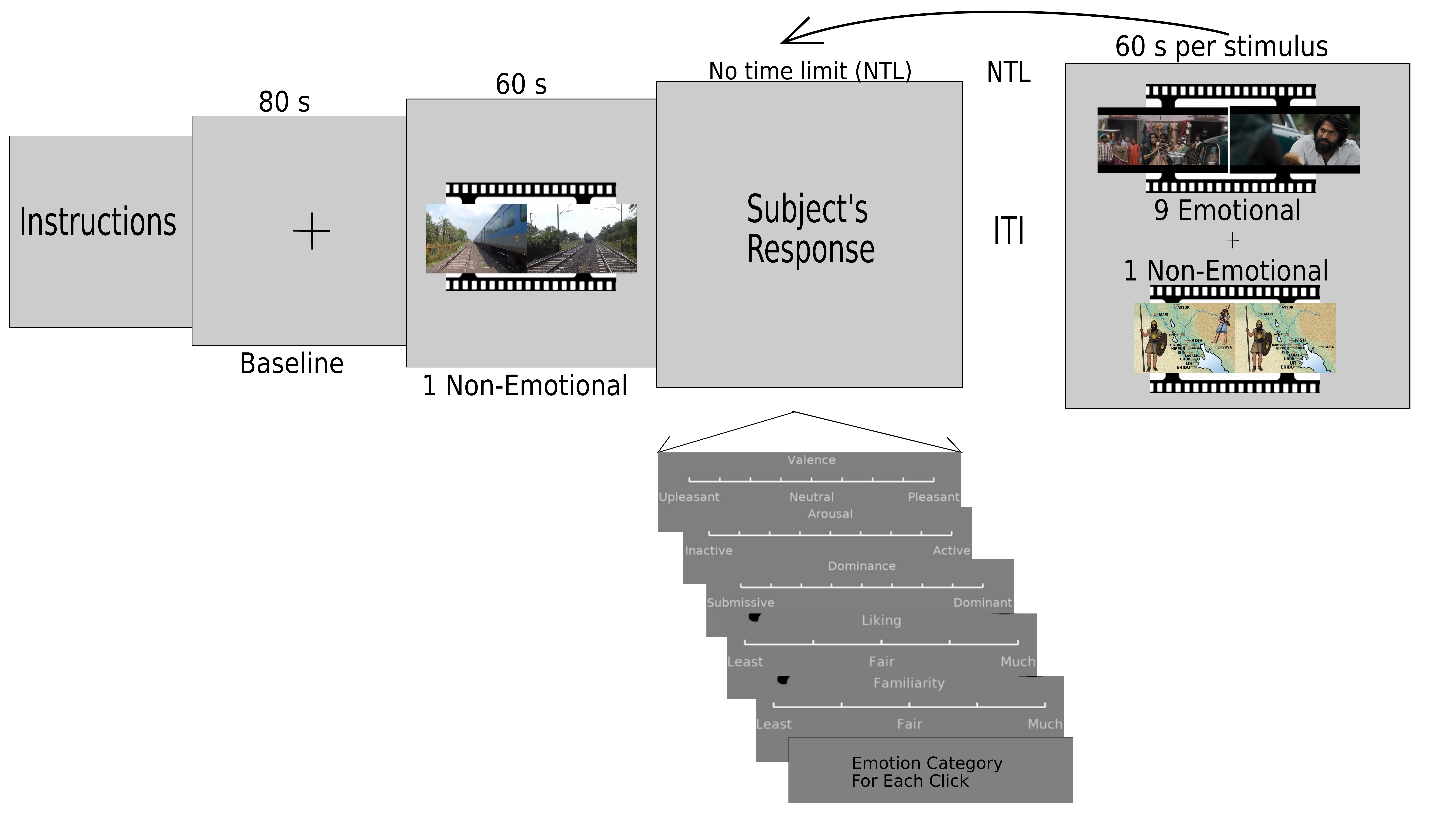}
    \caption{Each participant was shown 11 60-second long video stimuli. One non-emotional stimulus was shown after the baseline recording. Nine emotional stimuli were presented with order randomized. Another non-emotional stimuli was shown after the fifth and before the eighth video. There was no time limit during the ratings and inter-trial interval. Participants could resume watching the next stimulus after clicking the left button on the~mouse during the inter-trial interval.}
    \label{fig:Exp_Paradigm}
\end{figure}

The experiment paradigm is shown in the figure-\ref{fig:Exp_Paradigm} (figure is adapted from \cite{mishra2022dynamic}). Multimedia video stimuli for emotion elicitation were selected from a validated stimuli dataset on the Indian population \cite{mishra2021affective}. Participants were given instructions and provided information about rating scales. An eye-closed resting state was recorded for 80 seconds, followed by the presentation of eleven stimuli (nine emotional and two non-emotional stimuli with 60 seconds each). The innovative part of the experiment was capturing the emotional events during the continuous recording of brain activity stimulated with naturalistic emotional film stimuli. These emotional events were marked by the subjects while watching the film stimulus. Participants were instructed as follows-"while watching the stimulus at any point of time when you feel an emotion, you have to respond by performing a left mouse click anywhere on the screen". After watching each stimulus, participants had to respond on six self-assessment scales, including valence, arousal, dominance, liking, familiarity and relevance, and they had to categorize each emotional event in one of the emotional categories. During the emotion category selection, participants were shown three frames extracted around the click duration to help them recall the context and felt emotional experience in that context. Data for 420 emotional events for 24 self-reported emotions were collected with 40 participants. 

\subsection{EEG Data Acquisition \& preprocessing}
In the experiment room, participant sat (on the chair with an armrest) at a distance of 2 meters from the screen with a resolution of 640x480. Other logistics were keyboard and mouse to respond and Sennheiser CX 180 Street II in-Ear Headphones for the audio. A 128 Channel Geodesic EEG System 400 was attached to iMAC via an amplifier. Net station software running on the iMAC recorded the raw electrophysiological signal (sampling rate-250 Hz).

\begin{figure}[]
\includegraphics[width=\textwidth, height=0.5\linewidth]{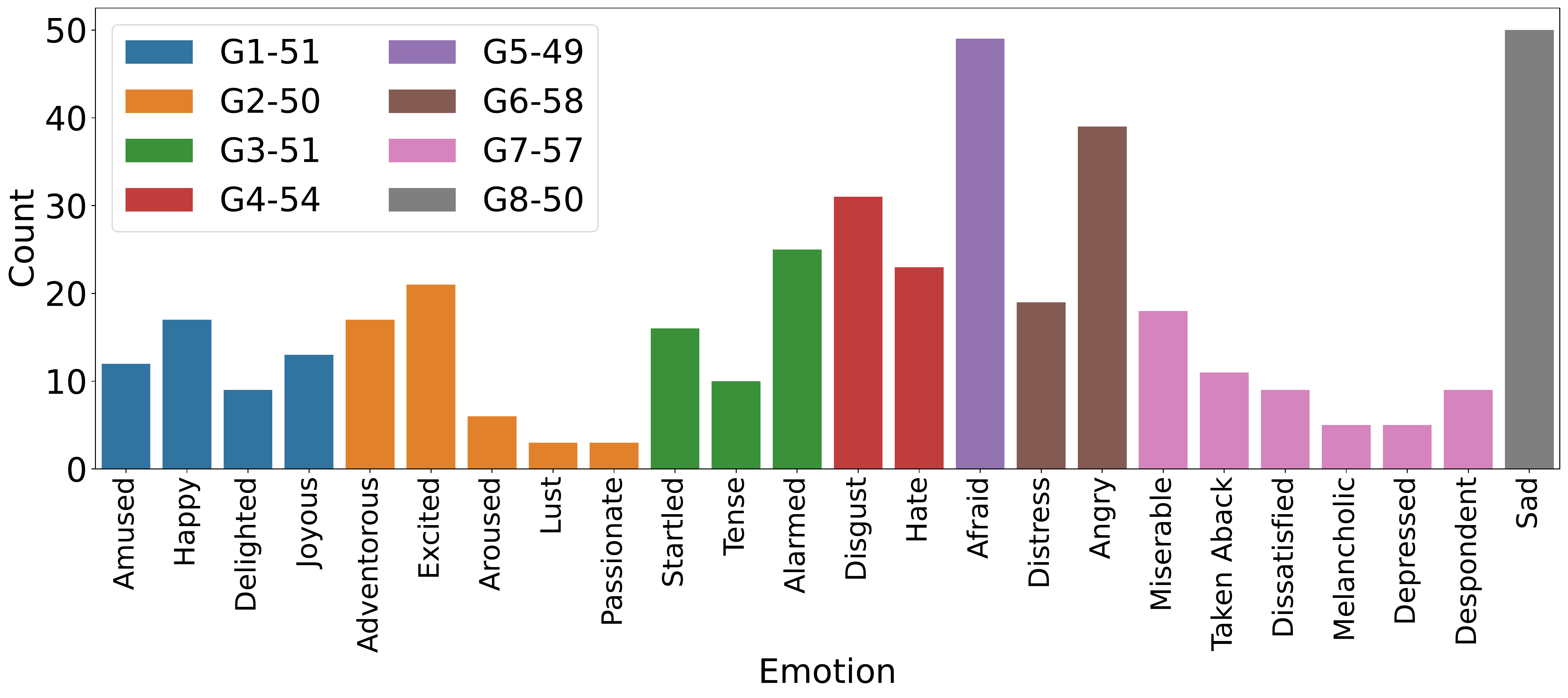}
\caption{ Representation and statistics of emotion groups: 
 The histogram plot shows the number of time participants has rated any emotion category to express their emotional experience. Histogram bars in the same color form an emotion group. The~legend shows the number of emotional instances per emotion group. The figure is recreated from \cite{mishra2022dynamic}. G1: happy, amused, delighted, joyous; G2: aroused, adventurous, excited, passionate, lust; G3: startled, tense, alarmed; G4: disgust, hate; G5: afraid; G6: distress, angry; G7: miserable, taken aback, dissatisfied, melancholic, depressed, despondent; G8: sad.
}
\label{fig:Grouping}
\end{figure}

A bandpass filter with passband 1Hz to 40Hz was applied to the raw EEG signal. EEG signal was also manually checked. Independent component analysis, followed by the ICLabel technique, was applied to detect components associated with noises, including eye blink, muscle noise, heart, line noise, and channel noise. The number of independent components varied from 25 to 45 across subjects (agreeing with the emotion analysis done by Hsu \cite{hsu2022unsupervised} on dense EEG with 128 channels). Emotional events, with seven seconds duration (six seconds before and one second after the click), were extracted from the preprocessed EEG signal. Considering seven seconds duration, we observed that 18 clicks by the participants were overlapping. These 18 emotional event related clicks were removed, and we had the final collection of 420 emotional events. These 420 emotional events are divided into eight emotion groups (shown in the figure-\ref{fig:Grouping}). The procedure to create these eight groups are described in \cite{mishra2022dynamic}). 

\subsection{Calculation of MS(s)}\label{micros}
\begin{figure}
    \centering
    \includegraphics[width=\linewidth, height=0.17\linewidth]{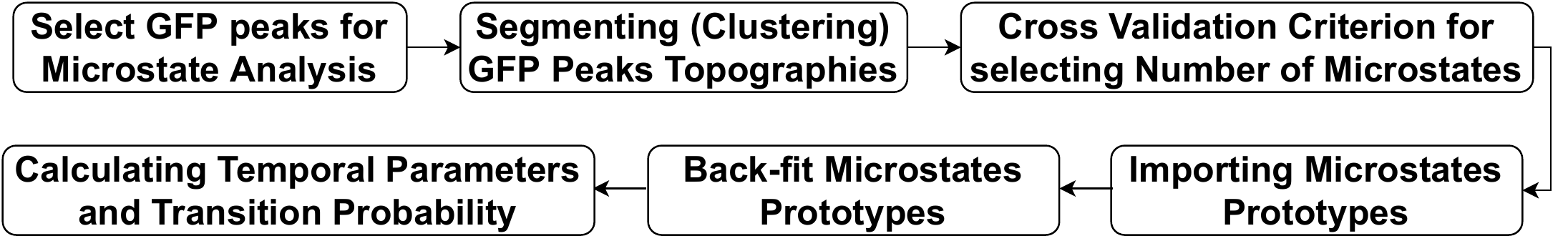}
    \caption{Methodology for segmenting EEG in MS(s)}
    \label{FIG:Methodology}
\end{figure}

The core consideration in the microstate analysis is to segment the EEG spatial representations into clusters with similar topographies using the clustering algorithms. The assumption is that the functional topographies that fall under the same cluster have similar underlying neural configurations and represent a functional microstate class. We have used the microstate eeglab toolbox for the MS calculation~\cite{poulsen2018microstate}. Figure~\ref{FIG:Methodology} shows the methodology adopted for the microstate analysis. The method to calculate the MS(s) is used with the parameters shown in the table-\ref{tab:multifractalStats}. 

\subsubsection{Global Field Power} 
The measure of global field power (GFP) corresponds to the spatial standard deviation, and it quantifies the variation of activity at each time point in the field, considering the data from all recording electrodes simultaneously. Further, only the time points with the GFP peaks are considered. In the series of GFP peaks, those that fulfil the parameters' criteria, including minimum peak distance, number of peaks, and GFP threshold (table-\ref{tab:multifractalStats}) are selected. The EEG maps at GFP peaks are concatenated across subjects. These maps are then given as input to the clustering algorithm. 

\subsubsection{Clustering EEG Maps} \label{clustering}
Modified k-means is used as the clustering algorithm. As the name implies, it is the modified version of k-means in two ways. First, it assumed that the topographical maps of the prototypical MS are polarity invariant (i.e. $a_k$ ~ $-a_k$). The second is that modified K-means models the activation of the MS(s), i.e. models the strength of the MS(s) for each time point. As stated earlier, the prototypical MS(s) are assumed to represent the underlying neural configuration. 

Microstate model: 

\vspace{-1cm}

\begin{flalign}
V_t = \Sigma_{k=1}^{N_\mu} {a_{kt}\Gamma_k}+E_t
\end{flalign}

\vspace{-0.8cm}

$V_t$ is modeling the topography at time point t using the MS(s) multiplied by the intensity of prototype plus residual noise. The objective of the modeling is the reduce the cost formalized in term of the following functional.

\vspace{-0.7cm}

\begin{flalign}
f = \frac{1}{N_T(N_s-1)}\Sigma_{t=1}^{N_T} ||V_t - \Sigma_{k=1}^{N_\mu} {a_{kt}\Gamma_k}||^2
\end{flalign}

\vspace{-0.7cm}

The minimum of the function (f) is obtained by assigning the MS with minimum distance to the sample at time t as follows \\

\vspace{-1cm}

\newcommand{\argmin}{\mathop{\mathrm{arg\,min}}}
\begin{flalign}
   d_{kt}^2 = V_t^{'}.V_t-(V_t^{'}.\Gamma_k)^2 \label{eq:6} \\
    \hat{L_t} = \underset{k}{\mathrm{argmin}}(d_{kt}^2) \label{eq:7} \\      
    \hat{a_{kt}} = V_t^{'}.\Gamma_k \label{eq:8} 
\end{flalign}

\vspace{-0.8cm}

where, $E_t$ is zero mean random noise, independent and identically distributed for all time instants. $\Gamma_k$ is the MS. $N_\mu,\; N_s,\; N_T,\; and \; V_t$ are model parameters defined as number of MS(s), number of electrodes, number of time points, and instantaneous topography, respectively. $\hat{L_t}$ is the label of the nearest (cluster) representative prototypical MS.



\textit{Number of repetitions:} Due to the stochastic nature of selecting the initial centroid, the k-means and modified version of it, even with the same settings and data, can give different results at each run. We repeated the clustering algorithm 50 times to address this problem and selected the best segmentation based on the CV criterion.  \\ \\
\textit{Stopping Criteria:} The clustering algorithm runs in iterations and assigns every new sample to one of the clusters. It keeps on repeating this procedure until the convergence criterion is achieved. In our case, we have used 1e-08 as the convergence criteria. As soon as the residual is below this threshold, the algorithm stops. We have used another criterion, a maximum number of iterations of 1000, to avoid the algorithm going in an infinite loop due to a stringent threshold.  \\ \\
\textit{Selecting the number of MS(s) clusters using measures of fit:} We have used the cross-validation criterion as a fitness measure to select the number of MS(s), which was introduced by Pascual \cite{pascual1995segmentation}. The low value of CV is better since it is related to the residual noise, $\epsilon$ as follows. \\
\begin{flalign}
CV = \hat{\sigma}^2.(\frac{N_s-1}{N_s-N_\mu-1})^2  \label{eq:9} \\
\hat{\sigma}^2 = \frac{\Sigma_{t=1}^{N_t}(V_t^{'}.V_t-(\Gamma_{kt}^{'}.V_t)^2)}{N_t(N_s-1)} \label{eq:10}
\end{flalign}

\vspace{-0.8cm}

where $\hat{\sigma}^2$ is an estimator of the variance of the residual noise. 

\begin{table}[H]
\begin{tabular}{cc}
\toprule
Parameters & Values\\
\midrule
Minimum peak distance & 10 \\
Number of peaks & 10000 \\
GFP Threshold & $\sigma_{GFPpeaks}$ \\
Clustering Algorithm & Modified K-means \\
Number of microstates & 2:8 \\
Number of repetitions & 50 \\
Maximum number of iterations & 1000 \\
Threshold & 1e-08 \\
Fitness or selection criteria & Cross Validation \\
Polarity & 0 \\
Temporal smoothing(TS) scheme & reject segments \\
Threshold time for MS redistribution during TS & 30 \\
\bottomrule
\end{tabular}
\caption{Parameters values used while calculating the EEG MS(s)}
\label{tab:multifractalStats}
\end{table}


\subsubsection{Backfitting EEG sequence with the MS(s)} \label{backfit}
We back-fitted the MS prototypes by replacing the scalp potential topography at any point with the nearest EEG MS topography. Since we had four MS prototypes, these prototypes are labelled with four labels say 1, 2, 3, and 4. Each observation is converted to a list of sequence of labels of these MS(s).


\subsection{Source localization of MS(s)}

In this study, we have used the sLORETA method for the pointwise distributed source localization \cite{pascual2002standardized} of preprocessed EEG signal. It is a distributed inverse imaging method. The current density estimate is based on the minimum norm ($l_2-norm$) solution, and localization inference is based on standardized values of the current density estimates. sLORETA is capable of exact (zero-error) localization. The objective function to be minimized to get zero error localization is $F = \| \Phi - \textbf{KJ} - c1 \|^2 + \alpha \|\textbf{J}\|^2$
where $\alpha \geq 0$ is a regularization parameter. This function is to be minimized with respect to \textbf{J} and c, for given \textbf{K}, $\Phi$ and $\alpha$. The explicit solution to this minimization problem is $\hat{\textbf{J}} = \textbf{T}\Phi$, where: $\textbf{T} = \textbf{K}^T\textbf{H}[HKK^TH + \alpha H]^+$ and $\textbf{H} = \textbf{I} - \textbf{11}^T/\textbf{1}^T\textbf{1}$ with $\textbf{H} \in \mathbb{R}^{N_E * N_E}$ denoting the centering matrix; $\textbf{I} \in \mathbb{R}^{N_E * N_E}$ the identity matrix; and $\textbf{1} \in \mathbb{R}^{N_E x N_E}$ is a vector of ones.

The labels were assigned to each time point of the EEG signal (as described in the section-\ref{backfit}). The estimated source regions on these time points are pooled as per the MS labels for each observation. It is assumed that the neural configuration during the time points assigned with the same labels will be more similar than the estimated neural configuration assigned to different labels. Then we calculated the mean source activity for each MS pool, which gave us the volumetric brain activity for each MS. The same procedure is applied to each emotional event observation to estimate the source corresponding to different MS(s). In this way, for each emotional and non-emotional and resting state condition, we obtained a set of estimated volumetric activity corresponding to MS(s). 

We checked the significance of estimated source activity using the bootstrapping method. We had a set of estimated MS(s) for each condition. The mean activity at each voxel was calculated, followed by the calculation of the maximum mean voxel activity. We calculated the sampling distribution by performing 5000 resampling with replacements from the population set. For each resample, we calculated the maximum mean voxel activity. Hence, the sampling distribution was created using the 5000 maximum mean voxel activity. The critical threshold is calculated by considering the Boneferroni corrected p-value. We rejected the null hypothesis if the original maximum mean activity is greater than the critical threshold. Furthermore, voxels with a value greater than the critical threshold are considered voxels with significant activity. 


\section{Results}

\subsection{EEG Microstates}
\begin{figure}[]
	\centering
		\includegraphics[width=0.55\textwidth, height=1.25\textwidth]{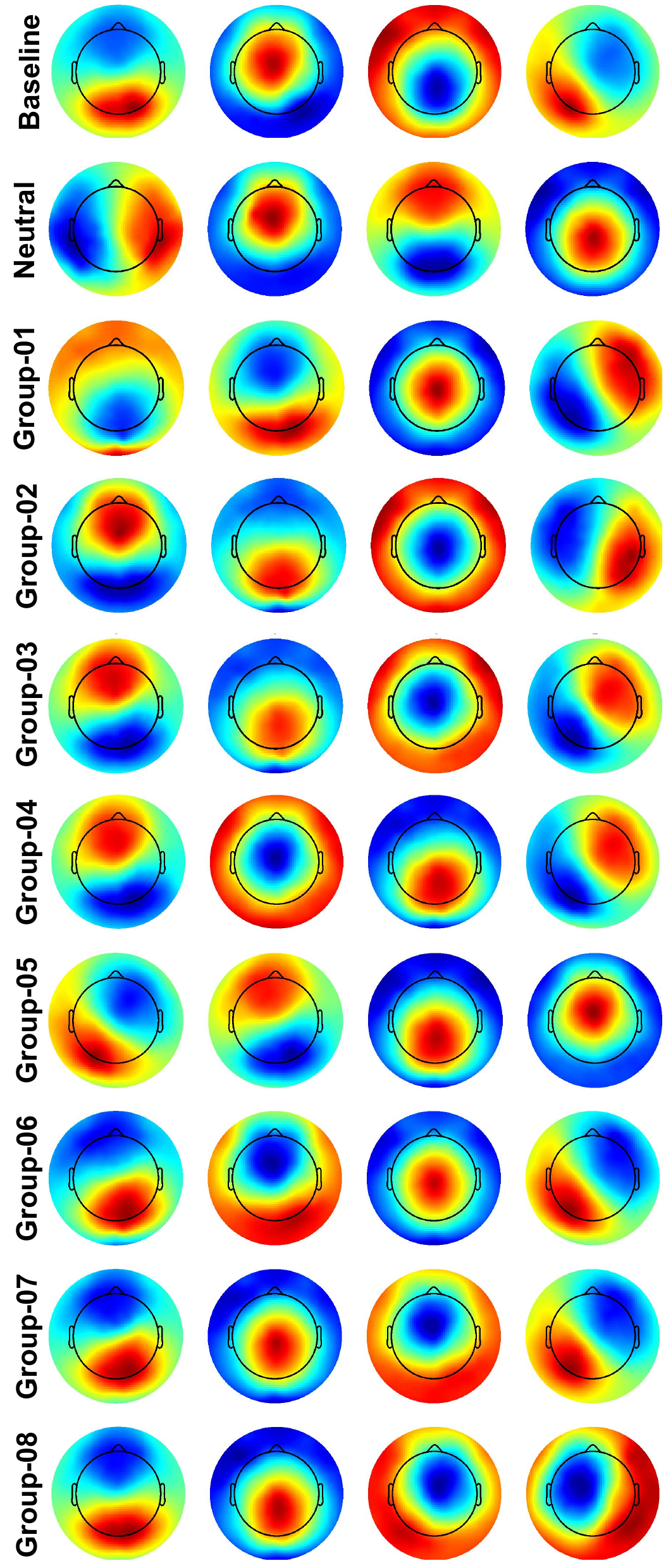}
	\caption{Microstate prototypes of different emotion groups for upper beta band. Microstate prototypes from left to right are arranged according the measure of fit criteria cross-validation (CV). Lower the value of CV is the better (method section-\ref{micros}), hence, from left to right the CV value is increasing in the plotted topographies. The topomaps for the eight emotional conditions are different from the non-emotion (or neutral) and baseline conditions (as shown in the distance table-\ref{tab:protodist})
	}
	\label{FIG:AllTopoUpperBeta}
\end{figure}

Using the microstate analysis, we calculated MS(s) from acquired EEG signals (as shown in fig~\ref{FIG:AllTopoUpperBeta}; see method section-\ref{micros}). We checked how different the MS(s) for all the emotional conditions are from the baseline and neutral conditions by calculating the distance between the MS(s). We observed that MS(s) for emotional conditions are significantly different (non-parametric permutation test with Bonferroni-Holm corrected p-value) from baseline and non-emotional (or neutral) conditions. The mean distance among MS(s) for eight emotion groups with baseline and non-emotional (or neutral) conditions is shown in the table-\ref{tab:protodist}. We observed significant dissimilarities between emotional conditions related topographies with the baseline and non-emotional (or neutral) condition-related topographies.  

In addition, we also tested the difference between MS(s) among emotion groups. We observed a significant difference in MS(s) for different emotional conditions. However, we did not correct p-values from multiple comparisons while comparing MS(s) among emotion groups. 

The difference in EEG-ms suggests that there may be functional cortical changes during the emotional experience, which is different from non-emotional and baseline conditions. This observation is based on the idea that EEG-ms represent spontaneous large-scale global neural coordination in time \cite{khanna2015microstates, michel2018eeg}. 



\begin{table}[H]
\begin{tabular}{cccccccccc}
& \textbf{Neutral} & \textbf{Group-01} & \textbf{Group-02} & \textbf{Group-03} & \textbf{Group-04} & \textbf{Group-05} & \textbf{Group-06} & \textbf{Group-07} & \textbf{Group-08} \\
\textbf{Baseline} & 0.66 & 0.45  & 0.61  & 0.46  & 0.27  & 0.56  & 0.32  & 0.51  & 0.64  \\
\textbf{Neutral}  & & 0.72  & 0.81  & 0.83  & 0.66  & 0.67  & 0.68  & 0.78  & 0.79  \\
\textbf{Group-01} & &  & 0.53  & 0.5   & 0.57  & 0.61  & 0.22  & 0.51  & 0.67  \\
\textbf{Group-02} & &  &  & 0.45  & 0.62  & 0.59  & 0.57  & 0.63  & 0.52  \\
\textbf{Group-03} & &  &  &  & 0.44  & 0.69  & 0.48  & 0.3   & 0.44  \\
\textbf{Group-04} & &  &  &  &  & 0.63  & 0.45  & 0.5   & 0.65  \\
\textbf{Group-05} & &  &  &  &  &  & 0.61  & 0.78  & 0.67  \\
\textbf{Group-06} & &  &  &  &  &  &  & 0.44  & 0.67  \\
\textbf{Group-07} & &  &  &  &  &  &  &  & 0.42 
\end{tabular}
\caption{\textbf{Distance between MS(s) for upper beta band:} The cell values in the table shows the mean distance between emotion groups' prototypes and baseline and non-emotional conditions.
}
\label{tab:protodist}
\end{table}

\begin{longtable}{cc|ccc}
\textbf{Group} & \textbf{Map} & \textbf{meanDiff} & \textbf{Stats} & \textbf{adjPval} \\ \hline
\endhead
\multicolumn{5}{c}{\textbf{Occurence: Emotion Group – Neutral}}    \\ \hline
Group-02& MS1 & 0.96& 6.68    & \textless{}0.001 \\
Group-04& MS1 & 0.77& 6.37    & \textless{}0.001 \\
Group-07& MS1 & 0.83& 6.92    & \textless{}0.001 \\
Group-08& MS1 & 1.1 & 8.19    & \textless{}0.001 \\
Group-01& MS2 & 0.48& 4.89    & 0.00504   \\
Group-01& MS1 & 0.8 & 4.86    & 0.00567   \\
Group-05& MS1 & 0.59& 4.8     & 0.00962   \\ \hline
\multicolumn{5}{c}{\textbf{Duration: Emotion Group – Neutral}}     \\ \hline
Group-03& MS1 & 20.46      & 6.35    & \textless{}0.001 \\
Group-07& MS1 & 19.1& 7.56    & \textless{}0.001 \\
Group-08& MS1 & 24.17      & 6.47    & \textless{}0.001 \\
Group-04& MS1 & 18.5& 5& 0.00319   \\
Group-02& MS3 & -15.92     & -4.55   & 0.0154    \\
Group-01& MS3 & -15.99     & -4.49   & 0.02106   \\
Group-01& MS2 & 10.49      & 4.37    & 0.03094   \\
Group-01& MS1 & 13.32      & 4.29    & 0.039     \\
Group-04& MS3 & -11.17     & -4.19   & 0.05016   \\ \hline
\multicolumn{5}{c}{\textbf{Coverage: Emotion Group – Neutral}}     \\ \hline
Group-01& MS2 & 0.07& 6.05    & \textless{}0.001 \\
Group-02& MS1 & 0.13& 7.7     & \textless{}0.001 \\
Group-03& MS1 & 0.13& 6.9     & \textless{}0.001 \\
Group-04& MS1 & 0.11& 6.41    & \textless{}0.001 \\
Group-06& MS1 & 0.1 & 6.1     & \textless{}0.001 \\
Group-07& MS1 & 0.12& 8.62    & \textless{}0.001 \\
Group-08& MS1 & 0.16& 8.62    & \textless{}0.001 \\
Group-01& MS1 & 0.11& 4.96    & 0.0035    \\
Group-07& MS3 & -0.1& -4.93   & 0.00408   \\
Group-08& MS3 & -0.1& -4.82   & 0.00966   \\
Group-02& MS3 & -0.09      & -4.55   & 0.01232   \\
Group-01& MS3 & -0.09      & -4.51   & 0.01491   \\
Group-04& MS3 & -0.07      & -4.34   & 0.0244 \\ \hline  
\caption{Comparison of Microstate statistics including occurrence, duration and coverage. \emph{p}-val is adjusted for the multiple comparisons using Bonferroni-Holm correction.
}
\label{tab:MicroStats}\\
\end{longtable}

\begin{figure}[]
\begin{subfigure}{\linewidth}
    \includegraphics[width=\linewidth, height=0.5\linewidth]{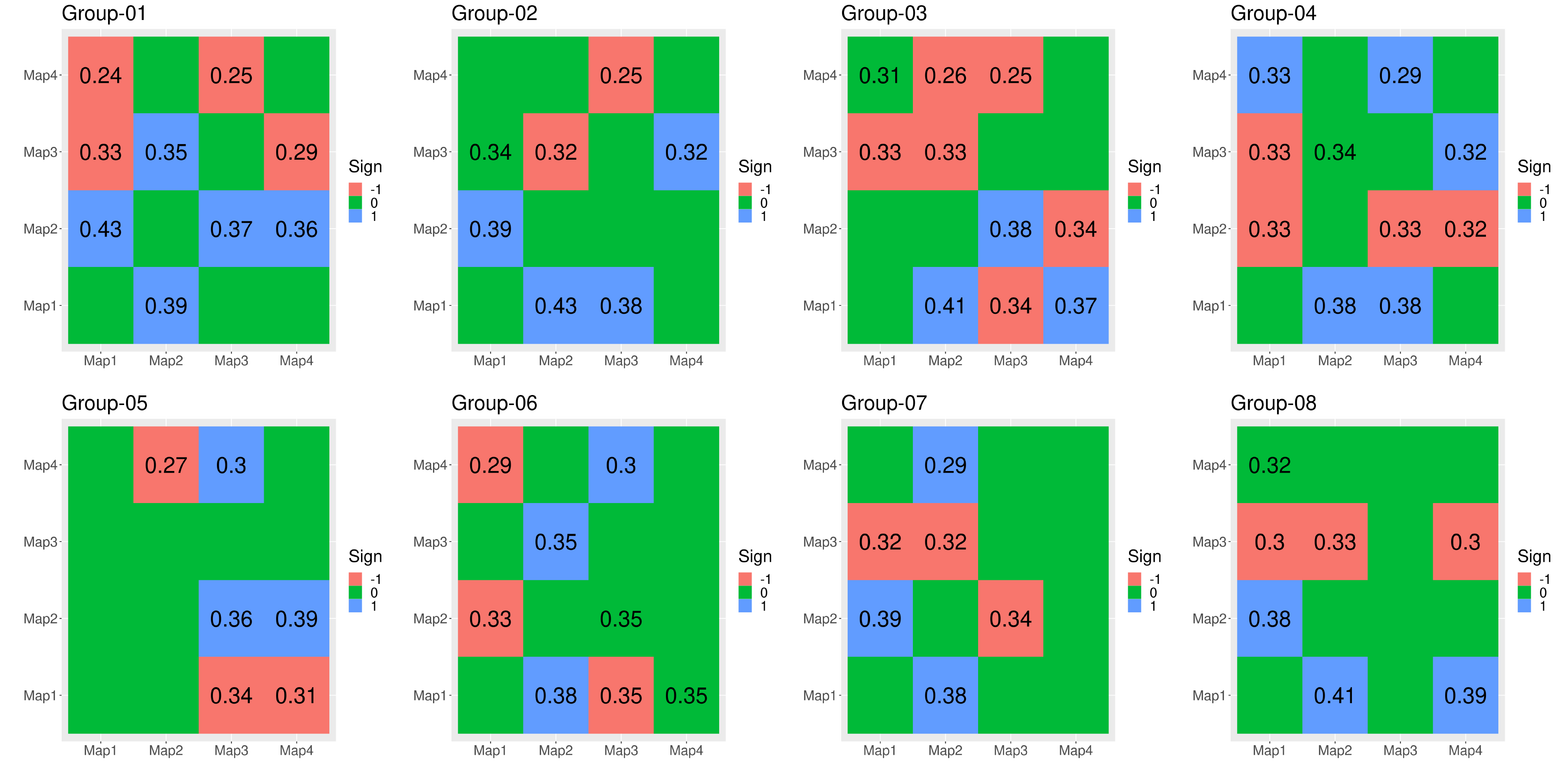}
    \caption{}
    \label{fig:MSStateTransBaseline} 
\end{subfigure}

\begin{subfigure}{\linewidth}
    \includegraphics[width=\linewidth, height=0.5\linewidth]{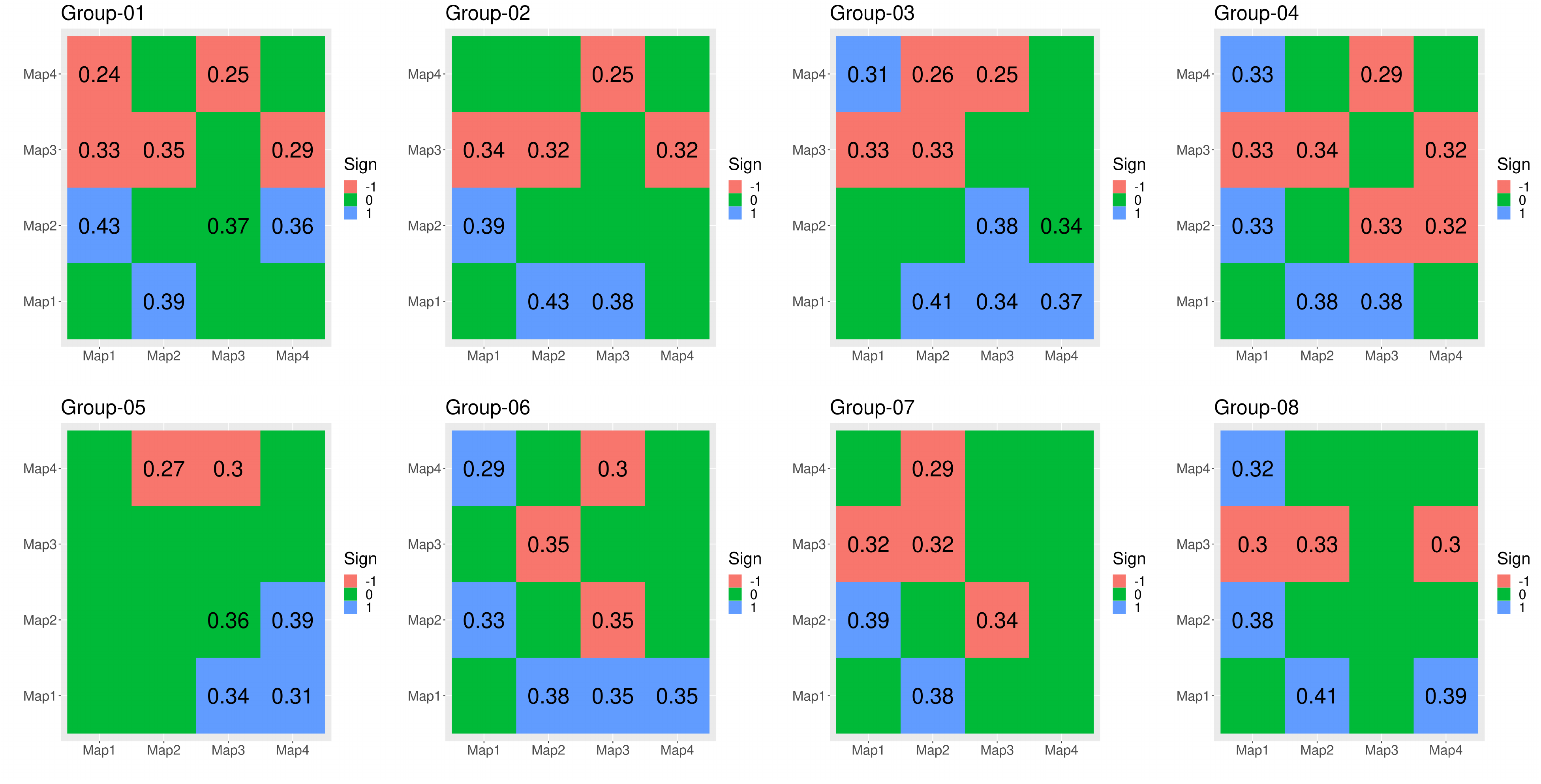}
    \caption{}
    \label{fig:MSStateTransNeutral} 
\end{subfigure}
\caption{(a) Significant transitions for different emotion conditions compared to baseline condition.(b) Significant transitions for different emotion conditions compared to neutral condition. 
}
\label{fig:TPMS}
\end{figure}

\subsection{Microstate Statistics}
\subsubsection{Occurrence}
Occurrence informs about the dominance of an MS in terms of the average number of times per second. We observed that MS1 occurred significantly more in emotion groups-1, 2, 4, 5, 7 and 8 than in non-emotional conditions. We did not observe any significant difference in MS occurrence between emotional groups and baseline conditions.

\subsubsection{Duration}
The duration parameter describes the average duration (in milliseconds) of a given MS. After staying in a particular MS for an average duration, the neural organization will be switched to another state. The duration of MS1 is higher for emotion groups-1, 3, 4, 7 and 8 than in neutral (or non-emotional) conditions. On the other hand, the duration of MS3 was greater for the non-emotional (or neutral) condition than for emotion groups. We did not observe any significant difference in MS duration between emotional groups and baseline conditions. When we calculated the difference in MS duration among different emotion groups for different MS(s), we observed that the duration of MS1 for emotion group-8 is greater than group-2, group-4 \& group-5 (table-\ref{tab:DurDiff}). In addition, the duration of MS1 for emotion groups-3, 4 and 7 is greater than group-5. In general, the duration of MS1 for group-5 is lesser than the duration of MS1 for four emotion groups, including group-3, 4, 7 \& 8. However, the duration of MS4 for the emotion group-5 is greater than emotion groups-2, 7 \& 8. During the sad emotional feeling, participants stay longer in MS1 than during the feeling of adventurous (group-2), hate (group-4) and afraid (group-8) emotions. 

\begin{longtable}{ccccccccc}
  & \textbf{G-1} & \textbf{G-2} & \textbf{G-3} & \textbf{G-4} & \textbf{G-5} & \textbf{G-6} & \textbf{G-7} & \textbf{G-8} \\ \hline
\endhead
\textbf{G-1} & 0 & 0 & 0 & 0 & 0 & 0 & 0 & 0 \\
\textbf{G-2} & 0 & 0 & 0 & 0 & 0 & 0 & 0 & 0 \\
\textbf{G-3} & 0 & 0 & 0 & 0 & 9.37 & 0 & 0 & 0 \\
\textbf{G-4} & 0 & 0 & 0 & 0 & 9.02 & 0 & 0 & 0 \\
\textbf{G-5} & 0 & 0 & 0 & 0 & 0 & 0 & 0 & 0 \\
\textbf{G-6} & 0 & 0 & 0 & 0 & 0 & 0 & 0 & 0 \\
\textbf{G-7} & 0 & 0 & 0 & 0 & 11.56& 0 & 0 & 0 \\
\textbf{G-8} & 0 & 8.88 & 0 & 9.44 & 12.42 & 0 & 0 & 0 \\ \hline
\caption{Difference in duration of MS1 among different emotion groups.}
\label{tab:DurDiff}\\
\end{longtable}

\subsubsection{Coverage}
The fraction of total time an MS is active is informed by the coverage parameter. We observed that MS1 is active for a longer duration during emotional conditions than in neutral condition. On the contrary, MS3 is active for a longer duration during the non-emotional (or neutral) condition. No significant difference is observed when coverage of MS for baseline conditions is compared with emotional conditions. 

\subsection{Transition between Microstates}
Transition probability from one MS to another can be interpreted as an encoded sequential activation of the neural assemblies that generate the MS(s). A comparison of transition values among MS(s) for emotional conditions with the neutral conditions shows that transition to MS3 and MS4 is more dominant for the neutral condition than for the emotional condition. On the contrary, the transition to MS1 and MS2 is more dominant in emotional conditions compared to the neutral condition (figure-\ref{fig:MSStateTransBaseline}).

For the high valence and high arousal emotional experience, transition to MS3 and MS4 is more in the baseline condition in comparison to emotional condition \ref{fig:MSStateTransNeutral}. On the contrary, the transition to MS1 and MS2 is more for the emotional condition than the baseline condition. A similar pattern persists for group-08, where a transition to MS3 is more for the baseline condition than the emotional condition. In addition, in comparison with the baseline (or resting state condition), transition to MS1 has a higher probability for emotional conditions, including group-1, group-2, group-4, group-7 and group-8. Whereas the transition to MS3 and MS4 is comparably more dominant in the baseline (or resting state) condition than it is for emotional groups, including group-1, group-2, group-3, and group-8. However, the transition to MS3 and MS4 is not as consistent as we have observed during the non-emotional (or neutral) condition. 

\subsection{Source Localization}
We performed source localization of MS(s). We observed that all four maps related to the non-emotional (or neutral) condition are localized to brain regions responsible for visual face and object recognition (for instance, FG) and higher order visual features (e.g., LG and OG) processing \ref{tab:LocalizedRegions}. In addition, regions contributing to attention functionality are also observed (including, MiFG). Moreover, regions providing the visual object and face recognition functionality are localized to MS1 (explicitly) for all emotional conditions. However, these regions are also observed in the list of estimated source regions for MS2, MS3 and MS4. Including visual object and face recognition related regions, MS3 and MS4 are localized to comprehension (e.g., L.TP, MiTG) and social-emotional processing (e.g., R.TP) related regions. The results described so far in this paragraph give some idea of MS-specific roles for neutral and emotional conditions. However, this overall specificity is not followed in the source estimation of maps for the baseline conditions. In the baseline (or resting condition), participants were instructed to sit stationary with their eyes closed. The maps for the baseline condition are source localized to comprehension, recognition, and socio-emotional processing regions. It shows the involvement of the domain-general system in both resting and emotional conditions. 


\begin{landscape}
\begin{longtable}{p{1.5cm}|p{5.15cm}|p{5.15cm}|p{5.15cm}|p{5.15cm}}
 & \textbf{MS1} & \textbf{MS2} & \textbf{MS3} & \textbf{MS4} \\ \hline
\endhead
\textbf{Baseline} & L.FG20, L.FG36, L.FG37, L.ITG20, L.MiTG38, L.STG38, R.FG20, R.FG36, R.FG37, R.ITG37, R.STG38 & L.FG20, L.FG36, L.FG37, L.ITG20, L.ITG37, L.MiTG38, L.STG38, R.FG20, R.FG36, R.FG37, R.ITG37, R.MiTG38, R.STG38   & L.FG20, L.FG36, L.FG37, L.IOG18, L.ITG20, L.ITG37, L.MiTG21, L.MiTG38, L.STG38, R.FG20, R.FG36, R.FG37, R.ITG37, R.MiTG38, R.STG38 & L.FG20, L.FG36, L.FG37, L.IOG18, L.ITG20, L.ITG37, L.MiTG21, L.MiTG38, L.STG38, R.FG20, R.FG36, R.FG37, R.ITG37, R.MiTG38, R.STG38 \\
\textbf{Neutral}  & L.FG20, L.LG18, L.MiFG10, L.MiFG11, R.C17, R.FG20, R.FG37, R.LG18, R.MiFG10, R.MiOG19 & L.FG20, L.IOG18, L.LG18, L.MiFG10, L.MiFG11, R.C17, R.FG20, R.FG37, R.LG18, R.MiFG10, R.MiOG19 & L.FG20, L.IOG18, L.LG18, L.MiFG10, L.MiFG11, L.MiOG18, L.SFG10, R.C17, R.FG20, R.FG37, R.IOG18, R.IOG19, R.LG18, R.MiFG10, R.MiOG18, R.MiOG19 & L.FG20, L.IOG18, L.LG18, L.MiFG10, L.MiFG11, L.MiOG18, L.SFG10, R.C17, R.FG20, R.FG37, R.IOG18, R.IOG19, R.LG18, R.MiFG10, R.MiOG18, R.MiOG19 \\
\textbf{Group-01} & L.FG20, R.FG20, R.FG36, R.FG37, R.ITG37 & L.FG20, R.FG20, R.FG36, R.FG37, R.ITG20, R.ITG37  & L.FG20, L.FG37, L.ITG20, L.MiTG38, R.FG20, R.FG36, R.FG37, R.ITG20, R.ITG37& L.FG20, L.FG36, L.FG37, L.ITG20, L.MiTG38, R.FG20, R.FG36, R.FG37, R.ITG20, R.ITG37  \\
\textbf{Group-02} & L.FG20, L.FG36, L.FG37, L.ITG20, L.ITG37, L.MiTG21, L.MiTG38, L.STG38, R.FG20, R.FG36, R.FG37, R.ITG20, R.ITG37 & L.FG20, L.FG36, L.FG37, L.ITG20, L.ITG37, L.MiTG38, L.STG38, R.FG20, R.FG36, R.FG37, R.ITG20, R.ITG37  & L.FG18, L.FG20, L.FG36, L.FG37, L.ITG20, L.ITG37, L.MiTG21, L.MiTG38, L.STG38, R.FG20, R.FG36, R.FG37, R.ITG20, R.ITG37 & L.FG18, L.FG20, L.FG36, L.FG37, L.ITG20, L.ITG37, L.MiTG21, L.MiTG38, L.STG38, R.FG20, R.FG36, R.FG37, R.ITG20, R.ITG37 \\
\textbf{Group-03} & L.FG20, L.FG37, R.FG20, R.FG36, R.FG37, R.ITG37 & L.FG20, L.FG36, L.FG37, L.ITG20, L.MiTG38, R.FG20, R.FG36, R.FG37, R.ITG37  & L.FG18, L.FG20, L.FG36, L.FG37, L.ITG20, L.ITG37, L.MiTG21, L.MiTG38, L.STG38, L.U38, R.FG20, R.FG36, R.FG37, R.ITG20, R.ITG37, R.MiTG38, R.STG38 & L.FG18, L.FG20, L.FG36, L.FG37, L.ITG20, L.ITG37, L.MiTG38, L.STG38, R.FG18, R.FG20, R.FG36, R.FG37, R.ITG20, R.ITG37, R.MiTG38, R.STG38   \\
\textbf{Group-04} & L.FG20, L.FG37, R.FG20, R.FG36, R.FG37, R.ITG37 & L.FG20, L.FG37, L.MiTG21, L.MiTG38, L.STG38, R.FG20, R.FG36, R.FG37, R.ITG20, R.ITG37, R.MiTG38, R.STG38   & L.FG20, L.FG37, L.ITG20, L.MiTG21, L.MiTG38, L.STG38, R.FG20, R.FG36, R.FG37, R.ITG20, R.ITG37, R.MiTG38, R.STG38& L.FG20, L.FG37, L.MiTG38, L.STG38, R.FG20, R.FG36, R.FG37, R.ITG20, R.ITG37, R.MiTG38, R.STG38\\
\textbf{Group-05} & L.FG20, L.FG37, L.IOG18, L.LG18, R.FG20 & L.FG20, L.FG37, L.IOG18, L.ITG37, L.LG18, L.MiOG18, R.FG20, R.FG37, R.LG18  & L.FG20, L.FG37, L.IOG18, L.ITG37, L.LG18, L.MiOG18, R.C17, R.FG20, R.FG37, R.LG18 & L.FG19, L.FG20, L.FG37, L.IOG18, L.ITG20, L.ITG37, L.LG18, L.MiOG18, R.C17, R.FG20, R.FG37, R.LG18   \\
\textbf{Group-06} & R.FG20, R.FG36, R.FG37, R.ITG37 & L.FG20, L.FG37, L.ITG20, L.MiTG38, L.STG38, R.FG20, R.FG36, R.FG37, R.ITG37, R.MiTG38, R.STG38 & L.FG20, L.FG37, L.ITG20, L.MiTG21, L.MiTG38, L.STG38, L.U38, R.FG20, R.FG36, R.FG37, R.ITG20, R.ITG37, R.MiTG21, R.MiTG38, R.STG38  & L.FG20, L.FG36, L.FG37, L.ITG20, L.MiTG21, L.MiTG38, L.STG38, L.U38, R.FG20, R.FG36, R.FG37, R.ITG20, R.ITG37, R.MiTG21, R.MiTG38, R.STG38 \\
\textbf{Group-07} & R.FG20, R.FG36, R.FG37, R.ITG37 & L.FG20, L.FG37, L.ITG20, L.MiTG21, L.MiTG38, L.STG38, R.FG20, R.FG36, R.FG37, R.ITG20, R.ITG37, R.MiTG38, R.STG38 & L.FG20, L.FG37, L.ITG20, L.MiTG21, L.MiTG38, L.STG38, L.U38, R.FG20, R.FG36, R.FG37, R.ITG20, R.ITG37, R.MiTG38, R.STG38& L.FG20, L.FG36, L.FG37, L.ITG20, L.MiTG21, L.MiTG38, L.STG38, L.U38, R.FG20, R.FG36, R.FG37, R.ITG20, R.ITG37, R.MiTG38, R.STG38   \\
\textbf{Group-08} & L.FG20, L.FG36, L.FG37, L.ITG20, L.ITG37, R.FG20, R.FG36, R.FG37   & L.FG20, L.FG36, L.FG37, L.ITG20, L.ITG37, L.MiTG38, L.STG38, R.FG20, R.FG36, R.FG37 & L.FG20, L.FG36, L.FG37, L.ITG20, L.ITG37, L.MiTG21, L.MiTG38, L.STG38, R.FG20, R.FG36, R.FG37, R.MiTG38, R.STG38& L.FG20, L.FG36, L.FG37, L.ITG20, L.ITG37, L.MiTG21, L.MiTG38, L.STG38, R.FG20, R.FG36, R.FG37, R.STG38  \\ \hline
\caption{Source Localization of EEG Microstate maps. LG: Lingual Gyrus, FG: Fusiform Gyrus, MiFG: Middle frontal gyrus, TP: Temporal pole, MiTG: Middle temporal gyrus, STG: Superior temporal gyrus, ITG: Inferior temporal gyrus, OG: Occipital gyrus.
}
\label{tab:LocalizedRegions}\\
\end{longtable}
\end{landscape}

\section{Discussion}

Based on our recent work \cite{mishra2022dynamic} emphasizing the importance of upper beta band in emotion related functional connectivity dynamics, in this work, we analyzed EEG signals in the upper beta band only. We segmented the EEG sequence into four MS(s) for each emotional, neutral (or non-emotional), and baseline conditions. The MS representations (shown in the figure-\ref{FIG:AllTopoUpperBeta}) for these conditions are different from each other as shown in the table-\ref{tab:protodist}. Compared to the baseline condition, MS maps for the non-emotional (or neutral) condition is more different from the emotional conditions. However, considering the transition dynamics in figure-\ref{fig:TPMS}, gives an impression that although the spatial configurations for emotion groups are relatively more similar to the baseline condition than spatial configurations for the non-emotional (or neutral) condition, the dynamics of transition of MS(s) for emotion groups is different from both baseline and non-emotional (or neutral) condition. It hints that similar domain generators might be involved in creating quasi-stable MS for baseline and emotional conditions, but the temporal transition between MS(s) is not the same. Moreover, the distance among MS(s) for different emotion groups are lesser than the distance of MS(s) for emotional groups with the MS(s) for neutral condition. The observed results align with the theory of constructed emotions proposed by \cite{barrett2017theory} that intrinsic or domain-general networks are involved in constructing emotional episodes. 

We observed that the duration of MS1 during the sad emotional feelings was more than during the feelings of adventurous (group-2), hate (group-4) and afraid (group-5) emotions. It might be that the elicitation of sad emotional feelings from an emotional stimulus needs more confirmatory evidence from the external social environment. Sad emotion is more pro-social in nature because it is associated with having pro-social goals and social expectations \cite{bastian2012feeling, bastian2015sad}. On the contrary, the afraid emotional feeling is a kind of evolutionary reflexive response which is more instinctive and demands a quick emotional response than confirmation \cite{rosenberg1990reflexivity}. Hence, the duration of MS4 for emotion group-5 is greater than emotion groups-2, 7 \& 8. It was pointed out \cite{jasper2018emotions} that reflex emotions are related to "fast time" compared to not so reflexive emotions, which are related to slow time. However, the evidence provided by us is very preliminary to support the claim by Jasper \cite{jasper2018emotions}. Hence, in future, more studies are needed to relate the average duration of MS(s) with different emotional feelings.

Comparison of the transition among MS(s) for the emotional conditions with the neutral (or non-emotional) condition showed that transition to MS3 during the neutral condition is more dominant than during the emotional experience. Similar results we observed in term of "coverage parameter" of MS(s)(table-\ref{tab:MicroStats}). MS3 of the neutral condition resembled the global neural coordination in the anterior posterior configuration. Generally, the anterior-posterior configuration is shown in the MS1 (ignoring the polarity) for the emotional condition. In addition, transition to MS1 is more for the emotional conditions than the non-emotional (or neutral) condition. The occurrence, duration and coverage parameters of MS(s) also confirm the dominant values for MS1. In our source localization results, we observed that MS1 for the emotional conditions is estimated to regions involved in visual features related to processing, including engagement of regions reported in object and face recognition. It might be that while watching the stimuli, subjects are transitioning to MS1 for the bottom-up feature level processing. So, MS1 is primarily representing the brain state for high-level visual feature processing, followed by MS2. 

On the other hand, MS-3 and MS-4 of emotional conditions involve brain regions reported during socio-emotional processing and semantic comprehension (for instance, bilateral temporal pole and middle temporal gyrus). We observed significant transitions to MS3 and MS4 during emotional conditions. According to the prevailing theory, the brain is not purely reflexive and stimulus-driven but consists of temporally structured intrinsic states. The metastable coordination dynamics allow the transition between these intrinsic states even during the rest \cite{friston1997transients, tognoli2014metastable}. However, the transition probability to MS3 and MS4 is less than the transition to MS1 and MS2. It might be that more evidence from the bottom-up mechanism is being collected while processing socio-emotional context in emotional multimedia stimuli. On the other hand, the transition probability to MS3 and MS4 is more for the non-emotional (or neutral) condition than the emotional condition. It could be because MS-3 and MS-4 for the non-emotional (or neutral) condition are localized to regions responsible for visual features related processing. 

In addition, there is a similarity in MS prototypes and results related to source localized estimations for resting state and different emotion groups (fig-\ref{FIG:AllTopoUpperBeta} \& table-\ref{tab:protodist}) in the upper beta band. These MS(s) have enough state-dependent information to emphasize the modality and content-specific mental process the state may represent \cite{michel2018eeg}. We interpret the approximate topographic similarity among resting-state activity and emotion-related activity in the framework of theories advocating domain-general brain representations for perception \cite{freeman2011dynamic,  freeman2020dynamic, barrett2017theory}. These domain-general networks, including default mode network, salience network, and frontoparietal control network, are reported in both resting-state analysis \cite{seeley2007dissociable}, and emotion analysis \cite{barrett2017theory, chen2018domain}. Functionally, the default mode network, salience network, and frontoparietal control network are involved in the representation of mental categories (or internal model), an adaptation of the internal model based on allostatically relevant salient error, and maintenance of the event representation, respectively \cite{barrett2017theory}. Despite having overlapping domain-general networks, it is to be noted that the long-range connectivity for these different emotion groups is distinguishable from each other \cite{mishra2022dynamic}. In addition, transition dynamics among MS(s) for resting-state and emotional conditions are different (figure-\ref{fig:MSStateTransBaseline}). It means that the domain-general systems dynamically interact with the environment to construct an emotional event. These domain-general systems facilitate the integration of conceptual knowledge about emotion categories into the perceptual process. 

In summary, for each emotional and non-emotional (baseline and neutral) condition, we calculated a set of MS topographies which alternate in discrete chunks with temporal duration. Metaphorically, if MS represents a potential neural configuration processing an atom of thought, can the temporal sequence and transition probability among these MS determine the content of the spontaneous emotional experience? During the neutral condition, we observed transition among MS states which are primarily localized to higher-order sensory feature processing regions, contrary to emotional experience-related MS, which additionally involves regions reported in socio-emotional processing. In addition, the comparison among transition probabilities led us to the interpretation that there is an interaction between neural configuration facilitating bottom-up feature processing with the neural configuration facilitating top-down higher order complex socio-emotional processing. Our results hint at the mechanism of emotional experience in which both bottom-up and top-down processing is involved in constructing an emotional experience. However, due to external multimedia stimulation, the transition to bottom-up processing states was higher than the transition to top-down processing states. On the other hand, during the neutral condition, the dominant neural configuration is related to sensory feature processing. In this work, we tried to understand the mechanism behind an emotional experience in terms of atomic neural configuration and transition among these different potential neural configurations. Since, to date, such study on healthy participants for different emotional conditions is not done, we encourage future research to do such kind of analysis in a large number of emotional conditions and in different cultural contexts to understand the mechanistic principle of emotional processing.




\bibliographystyle{unsrt}
\bibliography{sample}

\begin{thebibliography}{10}

\bibitem{von2016analytical}
Frederic von Wegner, Enzo Tagliazucchi, Verena Brodbeck, and Helmut Laufs.
\newblock Analytical and empirical fluctuation functions of the eeg microstate
  random walk-short-range vs. long-range correlations.
\newblock {\em Neuroimage}, 141:442--451, 2016.

\bibitem{poulsen2018microstate}
Andreas~Trier Poulsen, Andreas Pedroni, Nicolas Langer, and Lars~Kai Hansen.
\newblock Microstate eeglab toolbox: An introductory guide.
\newblock {\em BioRxiv}, (289850), 2018.

\bibitem{lehmann1998brain}
Dietrich Lehmann, WK~Strik, B~Henggeler, Thomas K{\"o}nig, and M~Koukkou.
\newblock Brain electric microstates and momentary conscious mind states as
  building blocks of spontaneous thinking: I. visual imagery and abstract
  thoughts.
\newblock {\em International Journal of Psychophysiology}, 29(1):1--11, 1998.

\bibitem{lehmann2010core}
Dietrich Lehmann, Roberto~D Pascual-Marqui, Werner~K Strik, and Thomas Koenig.
\newblock Core networks for visual-concrete and abstract thought content: a
  brain electric microstate analysis.
\newblock {\em Neuroimage}, 49(1):1073--1079, 2010.

\bibitem{dabiological}
F~Lopes da~Silva.
\newblock Biological aspects of eeg and magnetoencephalogram generation.
\newblock {\em Electroencephalography: Basic Principles, Clinical Applications
  and Related Fields}.

\bibitem{duc2019microstate}
Nguyen~Thanh Duc and Boreom Lee.
\newblock Microstate functional connectivity in eeg cognitive tasks revealed by
  a multivariate gaussian hidden markov model with phase locking value.
\newblock {\em Journal of neural engineering}, 16(2):026033, 2019.

\bibitem{zanesco2020within}
Anthony~P Zanesco, Brandon~G King, Alea~C Skwara, and Clifford~D Saron.
\newblock Within and between-person correlates of the temporal dynamics of
  resting eeg microstates.
\newblock {\em NeuroImage}, 211:116631, 2020.

\bibitem{croce2021rtms}
Pierpaolo Croce, Spadone Sara, Filippo Zappasodi, Antonello Baldassarre, and
  Paolo Capotosto.
\newblock rtms affects eeg microstates dynamic during evoked activity.
\newblock {\em Cortex}, 2021.

\bibitem{soni2018hyperactivation}
Sunaina Soni, Suriya~Prakash Muthukrishnan, Mamta Sood, Simran Kaur, and Ratna
  Sharma.
\newblock Hyperactivation of left inferior parietal lobule and left temporal
  gyri shortens resting eeg microstate in schizophrenia.
\newblock {\em Schizophrenia research}, 201:204--207, 2018.

\bibitem{rajagopalan2018machine}
Shyam~Sundar Rajagopalan, Sujas Bhardwaj, Rajanikant Panda, Venkateswara~Reddy
  Reddam, Chaitanya Ganne, Raghavendra Kenchaiah, Ravindranadh~C Mundlamuri,
  Thennarasu Kandavel, Kaushik~K Majumdar, Satishchandra Parthasarathy, et~al.
\newblock Machine learning detects eeg microstate alterations in patients
  living with temporal lobe epilepsy.
\newblock {\em Seizure}, 61:8--13, 2018.

\bibitem{baradits2020multivariate}
M{\'a}t{\'e} Baradits, Istv{\'a}n Bitter, and P{\'a}l Czobor.
\newblock Multivariate patterns of eeg microstate parameters and their role in
  the discrimination of patients with schizophrenia from healthy controls.
\newblock {\em Psychiatry research}, 288:112938, 2020.

\bibitem{de2020eeg}
Renate de~Bock, Amatya~J Mackintosh, Franziska Maier, Stefan Borgwardt, Anita
  Riecher-R{\"o}ssler, and Christina Andreou.
\newblock Eeg microstates as biomarker for psychosis in ultra-high-risk
  patients.
\newblock {\em Translational Psychiatry}, 10(1):1--9, 2020.

\bibitem{kalburgi2020children}
Sahana~Nagabhushan Kalburgi, Allison~P Whitten, Alexandra~P Key, and James~W
  Bodfish.
\newblock Children with autism produce a unique pattern of eeg microstates
  during an eyes closed resting-state condition.
\newblock {\em Frontiers in human neuroscience}, 14, 2020.

\bibitem{pal2020study}
Anita Pal, Madhuri Behari, Vinay Goyal, and Ratna Sharma.
\newblock Study of eeg microstates in parkinson’s disease: a potential
  biomarker?
\newblock {\em Cognitive Neurodynamics}, pages 1--9, 2020.

\bibitem{van2010eeg}
Dimitri Van~de Ville, Juliane Britz, and Christoph~M Michel.
\newblock Eeg microstate sequences in healthy humans at rest reveal scale-free
  dynamics.
\newblock {\em Proceedings of the National Academy of Sciences},
  107(42):18179--18184, 2010.

\bibitem{khanna2015microstates}
Arjun Khanna, Alvaro Pascual-Leone, Christoph~M Michel, and Faranak Farzan.
\newblock Microstates in resting-state eeg: current status and future
  directions.
\newblock {\em Neuroscience \& Biobehavioral Reviews}, 49:105--113, 2015.

\bibitem{poskanzer2020using}
Craig Poskanzer, Dan Denis, Ashley Herrick, and Robert Stickgold.
\newblock Using eeg microstates to examine post-encoding quiet rest and
  subsequent word-pair memory.
\newblock {\em BioRxiv}, 2020.

\bibitem{laganaro2017inter}
Marina Laganaro.
\newblock Inter-study and inter-individual consistency and variability of
  eeg/erp microstate sequences in referential word production.
\newblock {\em Brain topography}, 30(6):785--796, 2017.

\bibitem{d2021auditory}
David~F D’Croz-Baron, Lucie Br{\'e}chet, Mary Baker, and Tanja Karp.
\newblock Auditory and visual tasks influence the temporal dynamics of eeg
  microstates during post-encoding rest.
\newblock {\em Brain Topography}, 34(1):19--28, 2021.

\bibitem{james1890principles}
William James, Frederick Burkhardt, Fredson Bowers, and Ignas~K Skrupskelis.
\newblock {\em The principles of psychology}, volume~1.
\newblock Macmillan London, 1890.

\bibitem{efron1970minimum}
Robert Efron.
\newblock The minimum duration of a perception.
\newblock {\em Neuropsychologia}, 8(1):57--63, 1970.

\bibitem{dehaene2003neuronal}
Stanislas Dehaene, Claire Sergent, and Jean-Pierre Changeux.
\newblock A neuronal network model linking subjective reports and objective
  physiological data during conscious perception.
\newblock {\em Proceedings of the National Academy of Sciences},
  100(14):8520--8525, 2003.

\bibitem{rabinovich2015dynamical}
Mikhail~I Rabinovich, Alan~N Simmons, and Pablo Varona.
\newblock Dynamical bridge between brain and mind.
\newblock {\em Trends in cognitive sciences}, 19(8):453--461, 2015.

\bibitem{fries2005mechanism}
Pascal Fries.
\newblock A mechanism for cognitive dynamics: neuronal communication through
  neuronal coherence.
\newblock {\em Trends in cognitive sciences}, 9(10):474--480, 2005.

\bibitem{fries2015rhythms}
Pascal Fries.
\newblock Rhythms for cognition: communication through coherence.
\newblock {\em Neuron}, 88(1):220--235, 2015.

\bibitem{tognoli2014metastable}
Emmanuelle Tognoli and JA~Scott Kelso.
\newblock The metastable brain.
\newblock {\em Neuron}, 81(1):35--48, 2014.

\bibitem{nishida2013eeg}
Keiichiro Nishida, Yosuke Morishima, Masafumi Yoshimura, Toshiaki Isotani,
  Satoshi Irisawa, Kay Jann, Thomas Dierks, Werner Strik, Toshihiko Kinoshita,
  and Thomas Koenig.
\newblock Eeg microstates associated with salience and frontoparietal networks
  in frontotemporal dementia, schizophrenia and alzheimer’s disease.
\newblock {\em Clinical neurophysiology}, 124(6):1106--1114, 2013.

\bibitem{drissi2016altered}
Natasha~M Drissi, Attila Szak{\'a}cs, Suzanne~T Witt, Anna Wretman, Martin
  Ulander, Henriettae St{\aa}hlbrandt, Niklas Darin, Tove Hallb{\"o}{\"o}k,
  Anne-Marie Landtblom, and Maria Engstr{\"o}m.
\newblock Altered brain microstate dynamics in adolescents with narcolepsy.
\newblock {\em Frontiers in human neuroscience}, 10:369, 2016.

\bibitem{kikuchi2011eeg}
Mitsuru Kikuchi, Thomas Koenig, Toshio Munesue, Akira Hanaoka, Werner Strik,
  Thomas Dierks, Yoshifumi Koshino, and Yoshio Minabe.
\newblock Eeg microstate analysis in drug-naive patients with panic disorder.
\newblock {\em PloS one}, 6(7):e22912, 2011.

\bibitem{gschwind2016fluctuations}
Markus Gschwind, Martin Hardmeier, Dimitri Van De~Ville, Miralena~I Tomescu,
  Iris-Katharina Penner, Yvonne Naegelin, Peter Fuhr, Christoph~M Michel, and
  Margitta Seeck.
\newblock Fluctuations of spontaneous eeg topographies predict disease state in
  relapsing-remitting multiple sclerosis.
\newblock {\em NeuroImage: Clinical}, 12:466--477, 2016.

\bibitem{gao2017altered}
Fei Gao, Huibin Jia, Xiangci Wu, Dongchuan Yu, and Yi~Feng.
\newblock Altered resting-state eeg microstate parameters and enhanced spatial
  complexity in male adolescent patients with mild spastic diplegia.
\newblock {\em Brain topography}, 30(2):233--244, 2017.

\bibitem{zappasodi2017prognostic}
Filippo Zappasodi, Pierpaolo Croce, Alessandro Giordani, Giovanni Assenza,
  Nadia~M Giannantoni, Paolo Profice, Giuseppe Granata, Paolo~M Rossini, and
  Franca Tecchio.
\newblock Prognostic value of eeg microstates in acute stroke.
\newblock {\em Brain topography}, 30(5):698--710, 2017.

\bibitem{mishra2022dynamic}
Sudhakar Mishra, Narayanan Srinivasan, and Uma~Shanker Tiwary.
\newblock Dynamic functional connectivity of emotion processing in beta band
  with naturalistic emotion stimuli.
\newblock {\em Brain sciences}, 12(8):1106, 2022.

\bibitem{american2013diagnostic}
American~Psychiatric Association, American~Psychiatric Association, et~al.
\newblock Diagnostic and statistical manual of mental disorders: Dsm-5.
\newblock {\em United States}, 2013.

\bibitem{mishra2021affective}
Sudhakar Mishra, Narayanan Srinivasan, and Uma~Shanker Tiwary.
\newblock Affective film dataset from india (afdi): Creation and validation
  with an indian sample.
\newblock 2021.

\bibitem{hsu2022unsupervised}
Sheng-Hsiou Hsu, Yayu Lin, Julie Onton, Tzyy-Ping Jung, and Scott Makeig.
\newblock Unsupervised learning of brain state dynamics during emotion
  imagination using high-density eeg.
\newblock {\em NeuroImage}, 249:118873, 2022.

\bibitem{pascual1995segmentation}
Roberto~D Pascual-Marqui, Christoph~M Michel, and Dietrich Lehmann.
\newblock Segmentation of brain electrical activity into microstates: model
  estimation and validation.
\newblock {\em IEEE Transactions on Biomedical Engineering}, 42(7):658--665,
  1995.

\bibitem{pascual2002standardized}
Roberto~Domingo Pascual-Marqui et~al.
\newblock Standardized low-resolution brain electromagnetic tomography
  (sloreta): technical details.
\newblock {\em Methods Find Exp Clin Pharmacol}, 24(Suppl D):5--12, 2002.

\bibitem{michel2018eeg}
Christoph~M Michel and Thomas Koenig.
\newblock Eeg microstates as a tool for studying the temporal dynamics of
  whole-brain neuronal networks: a review.
\newblock {\em Neuroimage}, 180:577--593, 2018.

\bibitem{barrett2017theory}
Lisa~Feldman Barrett.
\newblock The theory of constructed emotion: an active inference account of
  interoception and categorization.
\newblock {\em Social cognitive and affective neuroscience}, 12(1):1--23, 2017.

\bibitem{bastian2012feeling}
Brock Bastian, Peter Kuppens, Matthew~J Hornsey, Joonha Park, Peter Koval, and
  Yukiko Uchida.
\newblock Feeling bad about being sad: the role of social expectancies in
  amplifying negative mood.
\newblock {\em Emotion}, 12(1):69, 2012.

\bibitem{bastian2015sad}
Brock Bastian, Peter Koval, Yasemin Erbas, Marlies Houben, Madeline Pe, and
  Peter Kuppens.
\newblock Sad and alone: Social expectancies for experiencing negative emotions
  are linked to feelings of loneliness.
\newblock {\em Social Psychological and Personality Science}, 6(5):496--503,
  2015.

\bibitem{rosenberg1990reflexivity}
Morris Rosenberg.
\newblock Reflexivity and emotions.
\newblock {\em Social Psychology Quarterly}, 53(1):3--12, 1990.

\bibitem{jasper2018emotions}
James~M Jasper.
\newblock {\em The emotions of protest}.
\newblock University of Chicago Press, 2018.

\bibitem{friston1997transients}
Karl~J Friston.
\newblock Transients, metastability, and neuronal dynamics.
\newblock {\em Neuroimage}, 5(2):164--171, 1997.

\bibitem{freeman2011dynamic}
Jonathan~B Freeman and Nalini Ambady.
\newblock A dynamic interactive theory of person construal.
\newblock {\em Psychological review}, 118(2):247, 2011.

\bibitem{freeman2020dynamic}
Jonathan~B Freeman, Ryan~M Stolier, and Jeffrey~A Brooks.
\newblock Dynamic interactive theory as a domain-general account of social
  perception.
\newblock In {\em Advances in Experimental Social Psychology}, volume~61, pages
  237--287. Elsevier, 2020.

\bibitem{seeley2007dissociable}
William~W Seeley, Vinod Menon, Alan~F Schatzberg, Jennifer Keller, Gary~H
  Glover, Heather Kenna, Allan~L Reiss, and Michael~D Greicius.
\newblock Dissociable intrinsic connectivity networks for salience processing
  and executive control.
\newblock {\em Journal of Neuroscience}, 27(9):2349--2356, 2007.

\bibitem{chen2018domain}
Taolin Chen, Benjamin Becker, Julia Camilleri, Li~Wang, Shuqi Yu, Simon~B
  Eickhoff, and Chunliang Feng.
\newblock A domain-general brain network underlying emotional and cognitive
  interference processing: evidence from coordinate-based and functional
  connectivity meta-analyses.
\newblock {\em Brain Structure and Function}, 223(8):3813--3840, 2018.

\end{thebibliography}

\section{Acknowledgements (not compulsory)}






\end{document}